\documentclass[prc,twocolumn,epsfig,nofootinbib,floatfix,showpacs]{revtex4}
\usepackage{graphics}
\usepackage{epsfig}
\usepackage{amsfonts}
\usepackage{amsmath}
\usepackage{bm}

\usepackage{color}

\begin{document}
\title{
Individual dipole toroidal states: main features and search in $(e,e')$ reaction
}
\author{V.O. Nesterenko$^{1,2,3}$, A. Repko$^4$, J. Kvasil$^5$, and P.-G. Reinhard$^6$ }
\affiliation{$^1$
Laboratory of Theoretical Physics,
Joint Institute for Nuclear Research, Dubna, Moscow
region, 141980, Russia}
\affiliation{$^2$
State University "Dubna", Dubna, Moscow Region, 141980, Russia}
\affiliation{$^3$
Moscow Institute of Physics and Technology,
Dolgoprudny, Moscow region, 141701, Russia}
\email{nester@theor.jinr.ru}
\affiliation{$^4$
Institute of Physics, Slovak Academy of Sciences, 84511, Bratislava, Slovakia}
\affiliation{$^5$
Institute of Particle and Nuclear Physics, Charles
University, CZ-18000, Praha 8, Czech Republic}
\affiliation{$^6$
Institut f\"ur Theoretische Physik II, Universit\"at Erlangen, D-91058, Erlangen, Germany}

\date{\today}

\begin{abstract}
Individual low-energy E1 toroidal and compressional states (TS and CS) produced
by the convective nuclear current ${\bf j}_{\rm c}$ were recently predicted for $^{24}$Mg
in the framework of quasiparticle random-phase-approximation (QRPA) with
Skyrme forces. In the present QRPA study with Skyrme parametrization SLy6,
we explore in more detail properties of these states (toroidal and
compressional responses, current distributions, and transitions probabilities
$B(E1K, 0^+0 \to 1^-K),\; B(E3K, 0^+0 \to 3^-K)$, $B(M2K, 0^+0 \to 2^-K)$ with
$K=$0 and 1) and analyze the possibility to discriminate and identify TS in inelastic
electron scattering to back angles. The interplay of the convective $\bold{j}_{\rm c}$
and magnetization  $\bold{j}_{\rm m}$ nuclear currents is thoroughly scrutinized.
A two-step scheme for identification of TS in $(e,e')$ reaction is proposed.
The key element of the scheme is the strong interference of the orbital and spin
contributions, resulting in specific features of E1 and M2 transversal form factors.
\end{abstract}

\maketitle
\section{Introduction}
\label{intro}
In our recent publications, individual low-energy E1 toroidal
and compressional states (TS and CS) in deformed nuclei $^{24}$Mg
\cite{Ne_PRL18} and $^{20}$Ne \cite{Ne_20Ne}  were predicted within the
quasiparticle random-phase-approximation
(QRPA) method with Skyrme forces.  In $^{24}$Mg, the TS is predicted to appear
as the lowest (E=7.92 MeV) dipole state with K=1 (where K is the projection of the
total angular momentum to the symmetry z-axis). Comparable
 individual low-energy dipole TS were found
for $^{10}$Be \cite{KE10Be_PRC17,KE10Be_arXiv19}, $^{12}$C
\cite{KE12C_PRC18}, and $^{16}$O \cite{KE16O_arXiv19}  using
the combined antisymmetrized molecular dynamics and generator coordinate
method \cite{KE_rew18}.
These predictions open a new promising path for the exploration
of vortical toroidal excitations.  Previously, the nuclear toroidal mode was
mainly studied as E1 isoscalar (T=0) toroidal giant resonance (TGR), see e.g.
\cite{Se81,Bas93,Mis06,Balb94,Ry02,Co00,Vr02,Pa07,Kv11,Rep13,Rei14,NePAN16,Rep17}
and references therein. However the experimental observation and identification
of the TGR is hampered by serious troubles. The resonance is usually masked by other
multipole modes (including dipole excitations of non-toroidal nature) located
in the same energy region. As a result, even the most relevant $(\alpha,\alpha')$
experimental data \cite{YoungexpZr,Uch04} still do not provide the direct evidence
for E1 TGR, see the discussion in Ref. \cite{Rep17}.
In this connection, individual low-energy E1 TS in light nuclei
have obvious advantages in exploration of the toroidal mode.
They are well separated from the neighbor dipole states and so can be easier
discriminated and identified in experiment than the TGR.

In this paper, we present a thorough exploration of various features of TS and CS
in $^{24}$Mg. In addition to TS at 7.92 MeV, the toroidal E1(K=1) excitation
at 9.97 MeV is analyzed.  The special attention
is paid to the impact of the magnetization nuclear current $\bold{j}_\mathrm{m}$ which,
being vortical, can affect the results for customary TS produced by the
convective current $\bold{j}_\mathrm{c}$.  It is found that $\bold{j}_\mathrm{m}$
can cause E1(K=0) TS
of predominant  magnetization origin. We also show that, similar to
recording of the vortical scissors \cite{sciss} and twist \cite{twist} modes
by strong orbital M1 and M2 transitions, the vortical TS in deformed nuclei
can be also signified by enhanced M2 transitions $0^+0 \to 2^-K$
between the ground state and $I^{\pi}K=2^-1$ rotational state based on
the TS with $I^{\pi}K=1^-1$.

It is known that there is a general, yet unresolved, problem how to search
and identify vortical nuclear states in experiment.
In this connection, we propose a two-step
scheme which could be useful in solution of this problem.
We apply this scheme to the search for the convective TS in $(e,e')$ scattering to backward
angles. Since E1 toroidal form factor is transversal  \cite{Bas93,Mis06},
this reaction looks most suitable.

At the first step of the scheme, the appropriate candidates for TS have to be chosen
from, e.g., QRPA calculations. These are
states with a significant toroidal E1 strength, clear toroidal distribution of the
nuclear current, and enhanced $B(M2)$ value. At the second step, the calculated transversal
E1 and M2 form factors for these states  are compared with experimental
$(e,e')$ data.
Our analysis shows that strong interference of the orbital and spin contributions
leads to specific features of E1 and M2 transversal form factors. As a result,
these form factors become very sensitive probes for the spin/orbital interplay.
So, if $(e,e')$ data  cannot be described by the spin contribution alone but are
well reproduced by spin + orbital contributions,
we may conclude that the orbital fraction is essential and correctly produced
by the calculations. Then we are confident that the
structure of chosen state is valid and its TS character and toroidal distribution
of the nuclear current may be considered as established.

The paper is organized as follows. In Sec. II, the calculation scheme
is outlined. In Sec. III, the numerical results are presented. The responses,
current fields, electromagnetic transitions, and E1 and M2 transversal
form factors are discussed in detail. In Sec. IV, the conclusions are drawn.

\section{Calculation scheme}
\label{sec-2}

The calculations for $^{24}$Mg are performed within the self-consistent QRPA
based on the Skyrme functional \cite{Ben03}. As in our earlier studies \cite{Ne_PRL18,Ne_20Ne},
we use the Skyrme parametrization SLy6 \cite{SLy6}. The QRPA code for axial
nuclei \cite{Repcode} exploits a 2D mesh in cylindrical coordinates.
The single-particle basis includes all the states from the bottom of the potential well
up to +55 MeV. The axial equilibrium deformation is $\beta$=0.536 as obtained by
minimization of the energy of the system. The volume pairing modeled by contact interaction
is treated at the BCS level \cite{Rep17}. The QRPA uses a large two-quasiparticle (2qp)
basis with the energies up to $\sim$ 100 MeV. The basis includes $\approx$ 1900
($K=0$) and $\approx$ 3600 (K=1) states. This basis guarantees that the Thomas-Reiche-Kuhn
sum rule \cite{Ring_book,Ne08} and isoscalar dipole energy-weighted
sum rule \cite{Harakeh_book_01} are exhausted by 100$\%$ and 97$\%$, respectively.

The toroidal and compressional modes  are  coupled \cite{Vr02,Pa07,Kv11}
and comparison of these vortical and irrotational patterns
of the nuclear flow is always instructive \cite{Kv11}.
So we inspect both vortical TS and irrotational CS.
The toroidal and compressional responses are quantified
in terms of reduced transition probabilities
\begin{equation}
\label{BE1Kalpha}
B_{\nu}(E1K, \alpha)=(2-\delta_{K,0})|\:\langle\nu|\:\hat{M}_{\alpha}(\;E1K)\:|0\rangle \:|^2
\end{equation}
where $|0\rangle$ and $|\nu\rangle$ mark the QRPA ground state and excited
$\nu$-th dipole state.
Matrix elements for the toroidal ($\alpha$=tor) and
compressional ($\alpha$=com)
transition operators  are \cite{Kv11,Rep13,Ne_PRL18}
\begin{eqnarray}
\label{TM_curl}
&&\langle\nu|\hat{M}_{\text{tor}}(E1K)|0\rangle
\\
&& = \frac{-1}{10 \sqrt{2}c}
\int d^3r r [r^2+d^s+d^a_K]
{\bf Y}_{11K} \cdot ( \bf{\nabla} \! \times \!
\delta {\bold j}^{\nu}({\bold r})) \; ,
\nonumber
\end{eqnarray}
\begin{eqnarray}
\label{CM_div}
&&\langle\nu|\hat{M}_{\text{com}}(E1K)|0\rangle
\\
&&=  \frac{-i}{10c}\int d^3r r[r^2+d^s-2d^a_K]
Y_{1K} (\bf{\nabla} \cdot \delta {\bold j}^{\nu}({\bold r})),
\nonumber
\end{eqnarray}
where
${\bf Y}_{11K}(\hat{\bf r})$ and $Y_{1K}(\hat{\bf r})$ are vector
and ordinary spherical harmonics; $\delta {\bold j}^{\nu}({\bold r})
=\langle \nu| \hat{\bold j}|0\rangle ({\bold r})$
is the current transition density (CTD);
$d^s= - 5/3 \langle r^2\rangle_0$ is the center-of-mass correction (c.m.c.)
in spherical nuclei \cite{Kv11,NVGS81,Rep19};
$d^a_K = \sqrt{4\pi/45}\langle r^2 Y_{20} \rangle_0 (3\delta_{K,0}-1)$
is the additional c.m.c. arising in axial deformed nuclei
\cite{Rep19,YNVG08}. The average values in c.m.c. are
$\langle f \rangle_0 = \int\:d^3r f \:\rho_0 /A$
where  $\rho_0$ is the g.s. density. As was checked,
these c.m.c. accurately remove spurious center-of-mass admixtures
in $^{24}$Mg.

The operator of the nuclear current
\begin{equation}
\hat{\bold j}(\bold r)= \hat{\bold j}_{\rm b}(\bold r) + \hat{\bold j}_{\rm cdt}(\bold r)
\end{equation}
includes the bare current $\hat{\bold j}_{\rm b}$ \cite{BMv1} and the
correction $\hat{\bold j}_{\rm cdt}$ \cite{RK19} taking into account
the effect of the current-dependent terms in the Skyrme functional.
The correction is necessary to recover the continuity equation
in Skyrme-QRPA calculations of the responses  and form factors.
The effect of
$\hat{\bold j}_{\rm cdt}$ is negligible in T=0 responses but can
be noticeable in T=1 and mixed cases \cite{RK19}.

The bare current consists of the convective and magnetization (spin) parts,
\begin{equation}\label{full_j}
 \hat{\bold j}_{\rm b}(\bold r)=
 \hat{\bold j}_{\rm c}(\bold r) + \hat{\bold j}_{\rm m}(\bold r)
= \frac{e\hbar}{m} \sum_{q =n,p}(\hat{\bold j}_{\;\rm c}^q(\bold r)
+ \hat{\bold j}_{\;\rm m}^q(\bold r)) \; ,
\end{equation}
where
\begin{eqnarray}
\hat{\bold j}^q_{\;\rm c}(\bold r)&=& -i\frac{e_{\text{eff}}^q}{2}
\sum_{k \epsilon q}(\delta({\bold r} - {\bold r}_k) {\bold \nabla}_k
+ {\bold \nabla}_k \delta({\bold r} - {\bold r}_k)) ,
\\
\hat{\bold j}^q_{\;\rm m}(\bold r)&=&
\frac{\bar{g}^q_{s}}{2} \sum_{k \epsilon q}
({\bold \nabla}_k \times \hat{\bold s}_{qk}) \delta({\bold r} - {\bold r}_k) .
\end{eqnarray}
Here $\hat{\bold s}_{qk}$ is the spin operator,
$e_{\text{eff}}^q$ are effective charges, $\bar{g}^q_{s}$ are spin g-factors,
$k$ numerates the nucleons. In the present calculations, we use the isoscalar
($e^{\rm n,p}_{\rm eff}=0.5$,
$\bar{g}^{\rm n,p}_s=(g_{s}^n + g_{s}^p) \eta /2 = 0.88 \eta$)
and proton ($e^{\rm p}_{\rm eff}=1, e^{\rm n}_{\rm eff}=0$,
$\bar{g}^{\rm n,p}_s=\eta g^{\rm n,p}_s$)
nuclear currents, where $g^{\rm p}_s = 5.58$ and
$g^{\rm n}_s = -3.82$ are bare g-factors and $\eta$ =0.7 is the quenching
\cite{Harakeh_book_01}.  The isoscalar current is relevant
for the comparison of the responses with data from isoscalar reactions like
$(\alpha,\alpha')$. The proton current is relevant for $(e,e')$ reaction.

The toroidal matrix element  (\ref{TM_curl})  with
$( \bf{\nabla} \! \times \! \delta {\bold j}^{\nu}({\bold r}))$
and compressional matrix element  (\ref{CM_div}) with
$(\bf{\nabla} \cdot \delta {\bold j}^{\nu}({\bold r}))$
are determined by the vortical and irrotational nuclear
flow, respectively.
The proton and neutron CTD from the
convective and magnetization parts of the nuclear current are
$\delta \bold {j}^q_{\;\rm c} = \langle \nu| \hat{\bold j}^q_{\;\rm c}|0\rangle$
and $\delta \bold{j}^q_{\;\rm m} =\langle \nu| \hat{\bold j}^q_{\;\rm m}|0\rangle$.

For magnetic quadrupole transitions $0^+0 \to 2^-K$,
the reduced transition probability is
\begin{equation}
B_{\nu} (M2K) = (2-\delta_{K,0})
|\:\langle\nu|\:\hat{M}(M2K)\:|0\rangle \:|^2
\end{equation}
with the transition operator
\begin{equation}
\hat{M}(M2K) = \mu_N\sum_{q=n,p} \sum_{k \epsilon q}
[ g_s^q \hat{\bold s}_{qk}
+ \frac{2}{3} g_l^q \hat{\bold l}_{qk}]
\cdot \bold{\nabla}_k [r^2 Y_{2K}]_k
\end{equation}
where $\hat{\bold l}_{qk}$ is the operator of the orbital moment,
the orbital g-factors are $g_l^q$=1 for protons and 0 for neutrons.
\begin{figure*} 
\centering
\includegraphics[width=10cm]{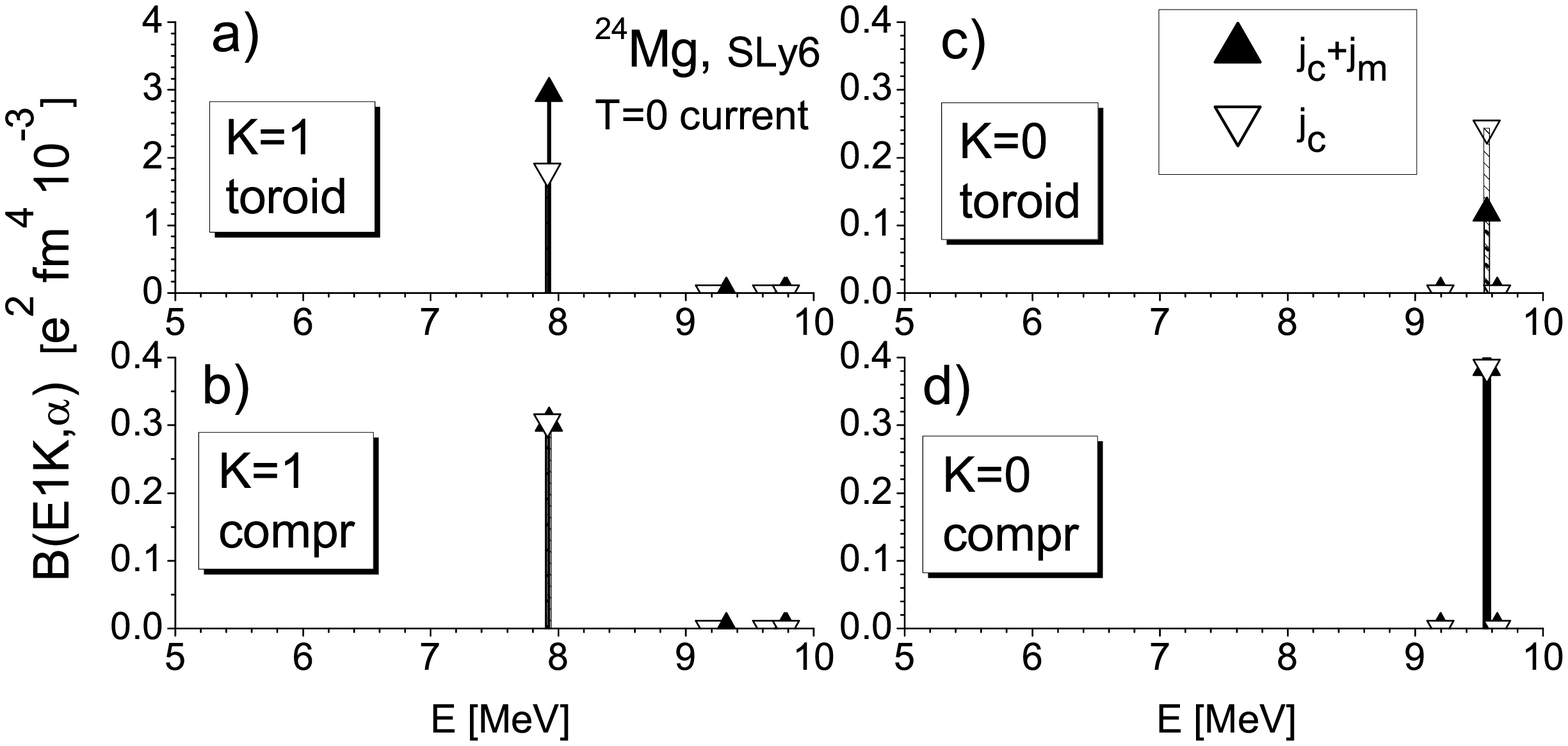}[h]
\caption{Toroidal (upper panels) and compressional (bottom panels)
B(E1K,$\alpha$)-strengths in $^{24}$Mg, calculated with T=0 nuclear current.
Calculations with (filled triangles) and
without (empty reverse triangles) $\bf{j}_{\rm m}$ are compared.}
\label{fig-1}
\end{figure*}

\section{Results and discussion}
\subsection{Toroidal and compression responses, current fields}

\begin{figure} 
\includegraphics[width=7.5cm,clip]{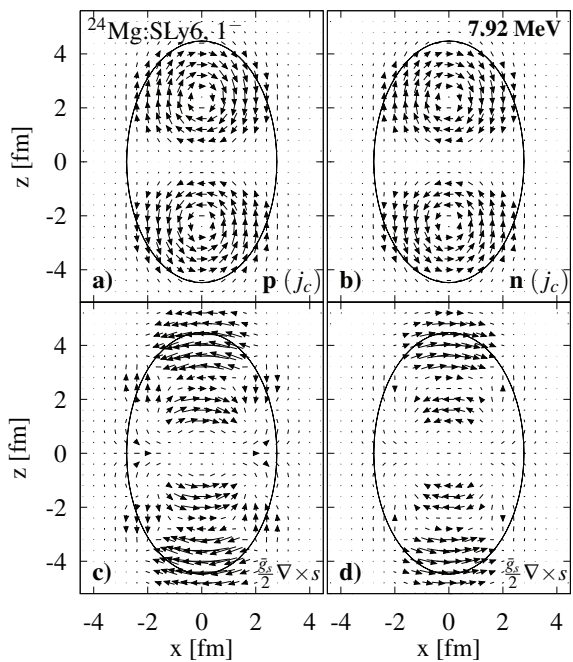}
\caption{QRPA proton (left) and neutron (right) fields of the convective
$\delta \bold {j}^q_{\rm c}$ (upper plots) and magnetic
$\delta \bold {j}^q_{\rm m}$ (bottom plots)
currents in the toroidal 7.92-MeV $K^{\pi}=1^-$ state.
In (c)-(d), the bare g-factors with the quenching are used.}
\label{fig-2}
\end{figure}
In Figure~\ref{fig-1}, the low-energy toroidal and compressional
transition strengths (\ref{BE1Kalpha}) in $^{24}$Mg are shown. They are calculated
with T=0 nuclear current relevant for isoscalar $(\alpha,\alpha')$ reaction.
The cases with and without $\bf{j}_{\rm m}$, are compared.
Plot (a) shows that only the K=1 state at 7.92 MeV exhibits the large
toroidal response. The toroidal nature of this state is additionally
confirmed by the proton and neutron  fields of the convective current,
shown in the plots (a)-(b) of Fig.~\ref{fig-2}.
Just this 7.92-MeV state was proposed  in \cite{Ne_PRL18} as
the individual low-energy TS. Due to the large axial quadrupole deformation
in $^{24}$Mg, the vortical flow of this state
is transformed from the familiar toroidal vortical ring  into the vortex-antivortex
dipole \cite{Ne_PRL18}. The 7.92-MeV state is not fully vortical since,
following  Fig.~\ref{fig-1} (b), it has
a small compressional irrotational response.
Even being small, the irrotational fraction can serve as a
{\it  doorway} for excitation of TS in various reactions. If
a reaction cannot  generate vortical excitations directly,
this can be done indirectly through the irrotational fraction.
\begin{figure} 
\includegraphics[width=7.5cm,clip]{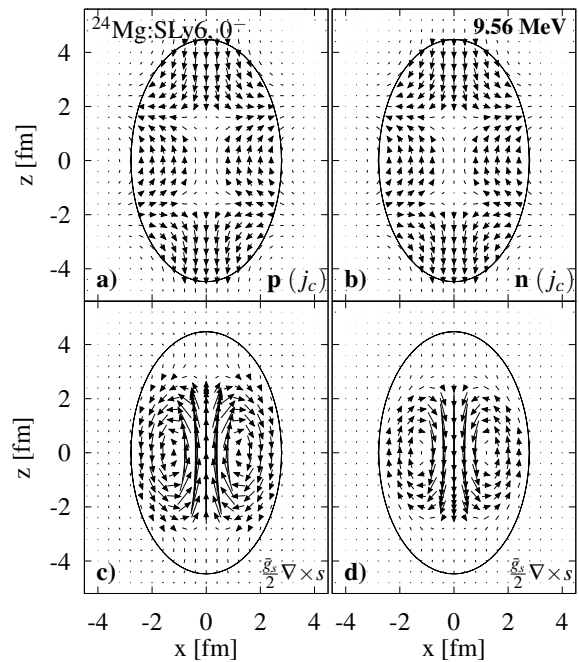}
\caption{As Fig. 2 but for the 9.56-MeV $K^{\pi}=0^-$ state.}
\label{fig-3}       
\end{figure}
\begin{figure*}[t] 
\includegraphics[width=10cm]{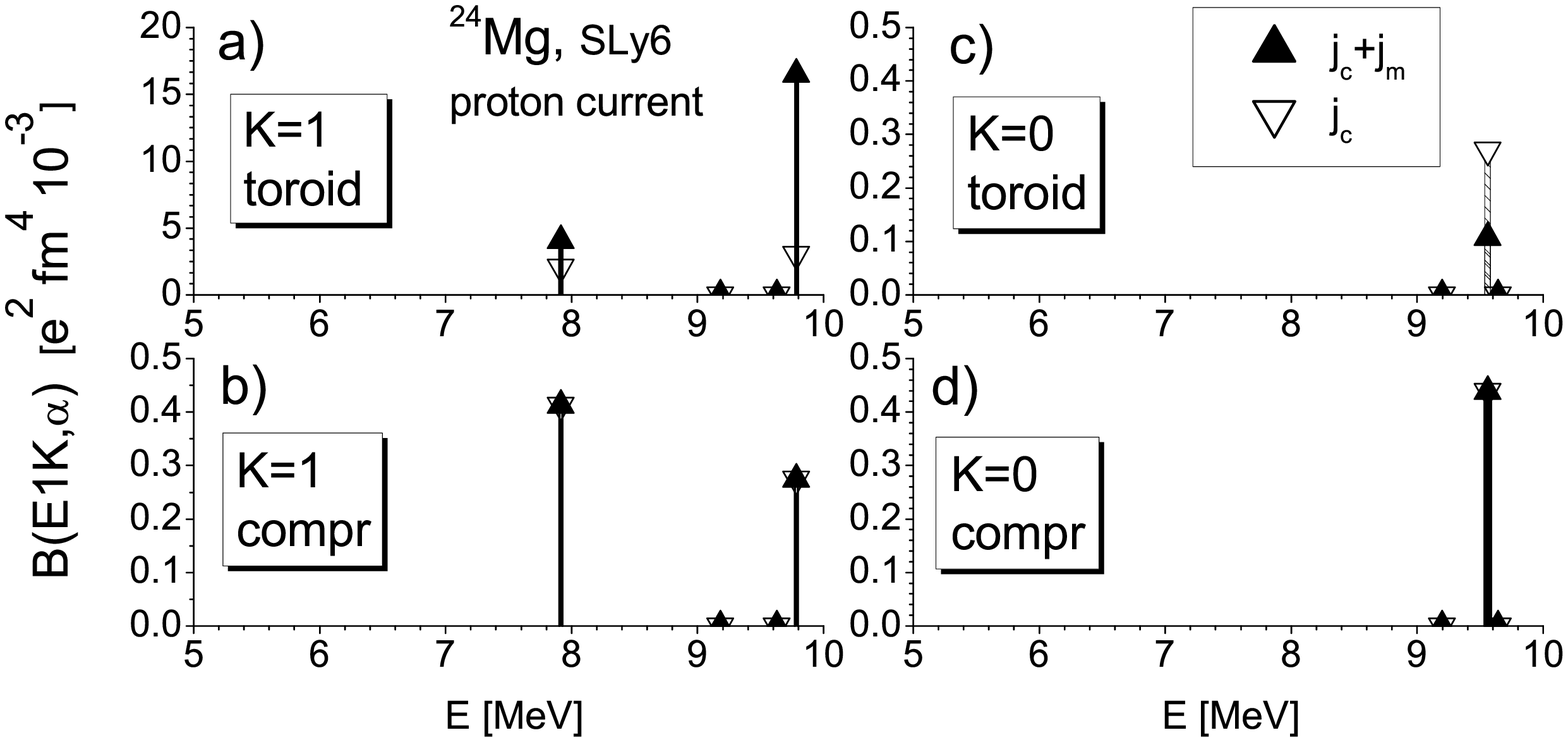}
\caption{As Fig. 1 but for the proton nuclear current,
see text for more detail.}
\label{fig-4}       
\end{figure*}

The plots Fig.~\ref{fig-1} (c)-(d) show that the compressional
strength exceeds the toroidal one for the K=0 state at 9.56 MeV. The
convective current $\delta \bold {j}^q_{\rm c}$ in this state (see
Fig.~\ref{fig-3} (a)-(b)) resembles the octupole flow for the $3^-$ state in
$^{208}$Pb \cite{RW87}.  This is not surprising since there is a strong
coupling between dipole and octupole modes in
nuclei with a large quadrupole deformation, like $^{24}$Mg. This coupling
should be especially strong in irrotational states like 9.56-MeV one.

\begin{figure}[h] 
\includegraphics[width=7.2cm,clip]{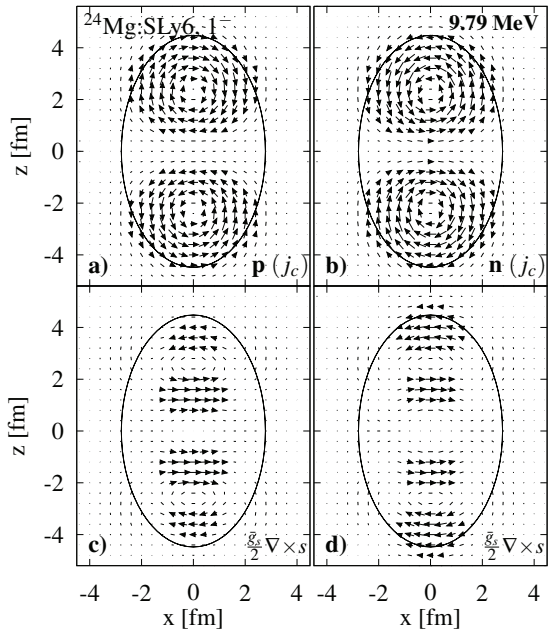}
\caption{As Fig. 2 but for 9.79-MeV $K^{\pi}=1^-$ state.}
\label{fig-5}       
\end{figure}

We now look at the impact of ${\bold j}_{\rm m}$. As seen from
Fig.~\ref{fig-1} (plots (b) and (d)),
the compressional strengths with and without $\bf{j}_{\rm m}$ are almost the same.
This is expected since vortical magnetization current should not affect
the irrotational compressional flow. At the same time,  plots (a) and (c) of Fig.~\ref{fig-1}
show that inclusion of ${\bold j}_{\rm m}$ significantly changes the toroidal strengths: it
is increased by $\sim 30\%$ in the K=1 7.92-MeV state and decreased to almost half
in the K=0 9.56-MeV state. Thus the impact of ${\bold j}_{\rm m}$ on the toroidal
strength is rather strong.

The proton and neutron magnetization current fields $\delta{\bold j}^q_{\rm m}$
for 7.92-MeV state shown in Fig.~\ref{fig-2} (c)-(d) are not toroidal. At the
same time, these fields  in the K=0 9.56-MeV state, shown in in  Fig.~\ref{fig-3}
(c)-(d), look like toroidal. Thus ${\bold j}_{\rm m}$, similar to ${\bold j}_{\rm c}$,
can cause a toroidal flow, which proves that {\it magnetization} vortical TS can exist.
\begin{table}
\caption{Main two-quasiparticle (2qp) components $ii'$
(denoted by Nilsson asymptotic quantum numbers $N n_z \Lambda$) in
some low-energy dipole $\nu$-states in $^{24}$Mg.
For each component, the forward amplitude $X^{\nu}_{ii'}$
and contribution $N^{\nu}_{ii'}$ to the state norm are listed.}
\label{tab-1}       
\begin{tabular}{lllll}
\hline
E [MeV] & K & main 2qp components & $X^{\nu}_{ii'}$ & $N^{\nu}_{ii'}$ \\\hline
7.92 & 1 & pp[211$\uparrow$-330$\uparrow$] & 0.73 & 0.54 \\
     &   & nn[211$\uparrow$-330$\uparrow$] & 0.62 & 0.39\\
9.56 & 0 & pp[211$\downarrow$-101$\downarrow$] & 0.62 & 0.39\\
     &   & nn[211$\downarrow$-101$\downarrow$] & 0.56 & 0.31\\
9.79 & 1 & nn[211$\uparrow$-330$\uparrow$] & -0.74& 0.55 \\
    &   & pp[211$\uparrow$-330$\uparrow$] & 0.65 & 0.43\\
9.93 & 0 &  pp[321$\uparrow$-211$\uparrow$] & -0.658 & 0.34\\
    &   & nn[321$\uparrow$-211$\uparrow$] & -0.50 & 0.25\\
\hline
\end{tabular}
\end{table}

Further, Fig.~\ref{fig-4} exhibits the toroidal and compressional strengths for
the effective charges ($e^{\rm p}_{\rm eff}=1, e^{\rm n}_{\rm eff}=0$,
$\bar{g}^{\rm p}_s=\eta g^{\rm p}_s$, $\bar{g}^{\rm n}_s=\eta g^{\rm n}_s$)
relevant for $(e,e')$ reaction. In this case,
the convective toroidal strength is determined only by the proton contribution.
Fig.~\ref{fig-4} (a) shows that  the convective toroidal strength
in 7.92-MeV state is similar to that in T=0 case and
comparable with the strength for the 9.79-MeV state which, following
Fig.~\ref{fig-5} (a)-(b), is also toroidal. Fig.~\ref{fig-4} (a) also shows
that, if ${\bold j}_{\rm m}$ is added,
then the toroidal response in 9.79-MeV state is significantly enhanced
and becomes dominant. The compressional responses are almost
not affected by ${\bold j}_{\rm m}$.

\begin{figure*} 
\includegraphics[width=12cm,height=6cm,clip]{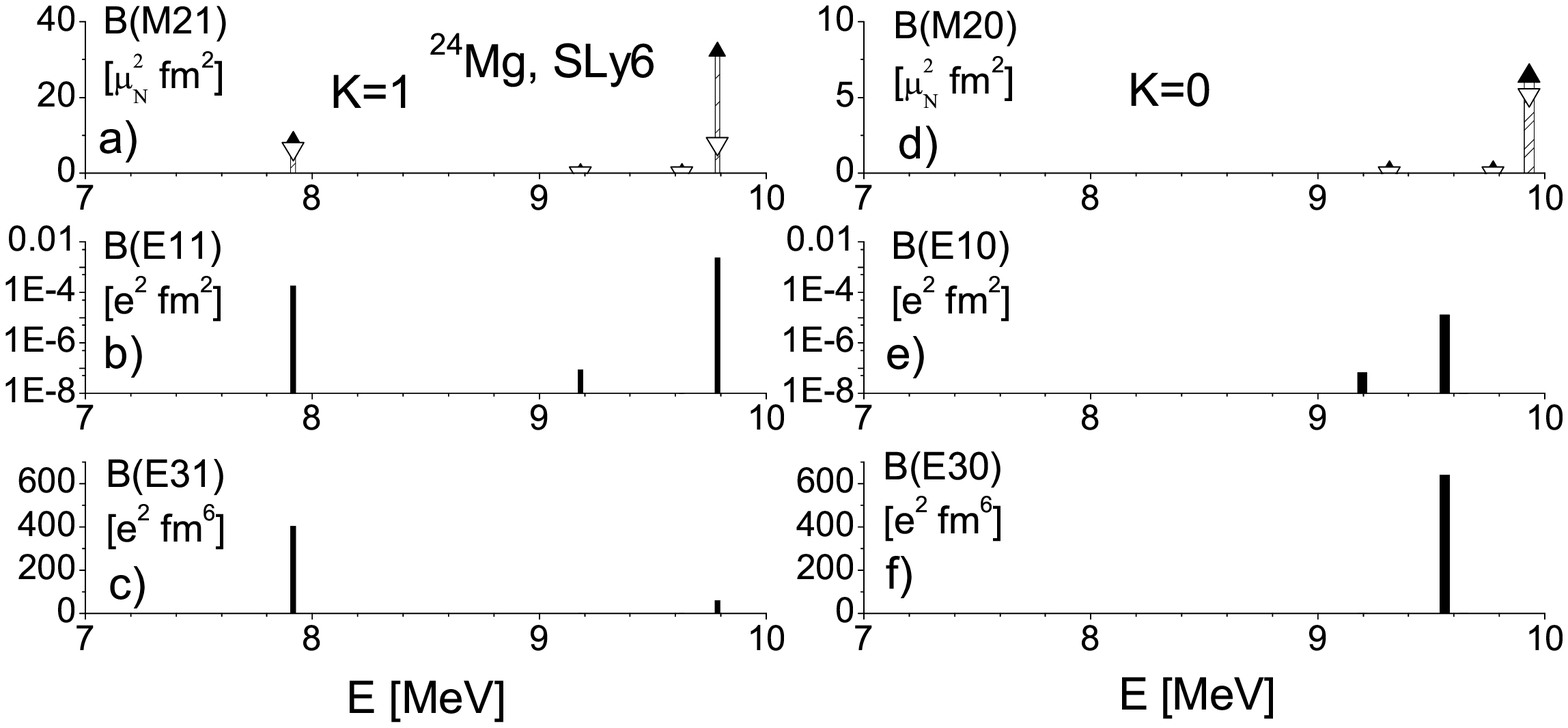}
\vspace{0.2cm}
\caption{$B(M2K, \; 0^+0_{\rm gs} \to 2^-K)$,
 $B(E1K, \; 0^+0_{\rm gs} \to 1^-K)$, and
$B(E3K,\; 0^+0_{\rm gs} \to 3^-K)$ values for
K=1 (left) and K=0 (right) in $^{24}$Mg.
In the plots (a) and (d), the total (filled triangles) and
orbital (empty reverse triangles) $B(M2K)$ values
are shown.
}
\label{fig-6}       
\end{figure*}

For a better understanding of these results, we provide in Table I more details on
the structure of dipole states discussed above. Besides, the structure of
K=0 state at 9.93 MeV is added since this state has a large $B(M2)$ value
to be discussed in the next subsection. Table I shows that K=1 states at 7.92
and 9.79 MeV are dominated by two (proton and neutron)
2qp components of almost the same weight. The toroidal response depends  on
the relative sign of $X^{\nu}_{ii'}$ in nn- and pp-components. In the 7.92-MeV state,
the proton and neutron $X^{\nu}_{ii'}$ have the same sign. As a result, proton
and neutron toroidal flows in Fig.~\ref{fig-2} are in phase
and we get for this state a large T=0 toroidal strength, see Fig.~\ref{fig-1} (a).
Instead, in the 9.79-MeV state, the proton and neutron amplitudes
$X^{\nu}_{ii'}$ have opposite signs. This makes the proton and neutron
toroidal flows in Fig.~\ref{fig-5} (a)-(b) also opposite.
The obtained destructive interference leads to the suppression of T=0 toroidal
strength in this state. Further, the different signs of the proton and neutron
$X^{\nu}_{ii'}$  in the 9.79-MeV state result in a significant enhancement of
the magnetization current (due to the constructive cooperation of the proton
and neutron g-factors). For this reason,
inclusion of ${\bold j}_{\rm m}$ leads a large increase of the total
toroidal vortical strength in this state, see Fig.~\ref{fig-4} (a). Furthermore,
since absolute values of the proton $X^{\nu}_{ii'}$ in
7.92-MeV and 9.79-MeV states are similar, the convective toroidal responses
for these states, shown  in Fig.~\ref{fig-4} (a),
are also comparable.

Note that the dominant 2qp components in Table I do not have spin-flip and so
favor the orbital vortical flow.
Altogether, the above analysis confirms the previous conclusions
\cite{Ne_PRL18,RW87,Ne_Dres} that toroidal flow in nuclei is mainly
determined  by the interplay of major 2qp components.

\subsection{Electromagnetic transitions}

For our aims, it is instructive to consider electromagnetic transitions from
the ground state to the rotational bands built on the toroidal and
compressional band heads. Below we inspect electric dipole
$B(E1K, \; 0^+0_{\rm gs} \to 1^-K)$, electric octupole
$B(E3K,\; 0^+0_{\rm gs} \to 3^-K)$, and magnetic quadrupole
$B(M2K, \; 0^+0_{\rm gs} \to 2^-K)$  reduced transition probabilities
with $K=0,1$.

As mentioned in the Introduction, the $B(M2K)$ value can be used
as an additional fingerprint of vortical toroidal states. Indeed, the
vortical scissors  \cite{sciss} and twist \cite{twist} modes are characterized
by enhanced orbital M1 and M2 transitions, respectively. Further,
the experimental techniques to extract M1 and M2 transition strengths in various reactions
are now available, see e.g. determination of M2 strength
from $(e,e')$ reaction \cite{PVNC99}. Last but not least,
the identification of mixed states by weak  E2 and large orbital M1 transitions
\cite{Pie08} shows that comparison of
electric and magnetic transitions is a useful identification tool.
Then it is worth to employ electromagnetic transitions for characterization of TS.
\begin{table} 
\caption{
The calculated reduced transition probabilities
$B(M2K, 0^+0_{\rm gs} \to 2^-K)$, $B(E1K, 0^+0_{\rm gs} \to 1^-K)$, and
$B(E3K, 0^+0_{\rm gs} \to 3^-K)$ for low-energy $\nu$-states
considered in Table I. For $B(M2K)$,
the total, spin, and orbital values are given.
The Weisskopf units \cite{BMv1} for $^{24}$Mg are used:
$B(M2)_{W.u} = $ 13.74 $\mu_N^2 {\rm fm}^2$,
$B(E1)_{W.u} = $ 0.537 $e^2 {\rm fm}^2$,
$B(E3)_{W.u} = $ 34.23 $e^2 {\rm fm}^6$.
}
\label{tab-2}       
\begin{tabular}{lllllll}
\hline
E & K & $B(M2)_{\rm tot}$ & $B(M2)_{\rm spin}$ & $B(M2)_{\rm orb}$
& $B(E1)$ & $B(E3)$
\\
MeV & & W.u. & W.u & W.u. & W.u. & W.u.
\\\hline
7.92 & 1 & 0.70 & 0.01 & 0.52 & 3.2 $\cdot 10^{-4}$& 12  \\
9.56 & 0 & -  & - & - &2.4 $\cdot 10^{-5}$ & 19\\
9.79 & 1 & 2.34 & 0.49 & 0.62 &4.2 $\cdot 10^{-3}$ & 1.7\\
9.93 & 0 & 0.93 & 0.01 & 0.75 & -               & - \\
\hline
\end{tabular}
\end{table}

The reduced transition probabilities $B(M2K)$,
$B(E1K)$,
 and  $B(E3K)$
in $^{24}$Mg are shown in Fig.~\ref{fig-6}. In panels (a) and (d), the total
and orbital ($g^q_s=0$) $B(M2K)$ strengths are compared. It is easy to see
that there is a remarkable correspondence between total/orbital $B(M21)$
in Fig.~\ref{fig-6} (a)  and total/orbital toroidal $B(E11)$
in Fig.~\ref{fig-4} (a). This proves that
$2^-1$ states based on the toroidal  K=1 band heads exhibit
large orbital $B(M21)$, i.e. there is a clear
correlation between toroidal E11 and orbital M21
strengths. So, for low-energy dipole states, an enhanced orbital $B(M21)$ values can be
used as an indicator for the toroidal mode.

Further, Table II shows that, in 7.92-MeV and 9.79-MeV states, orbital
$B(M21)$ strengths reach 0.52  and 0.62 W.u., i.e. are rather large.
In both states, the orbital strength dominates over the spin one,
especially  in 7.92-MeV state. Instead, the E11 strength in these states
is $\sim 10^{-3}-10^{-4}$
W.u., i.e. very weak.  This situation is similar to that for mixed-symmetry
states with its enhanced M1 and weaken E2 transitions \cite{Pie08}
(with the difference that mixed-symmetry states
are mainly isovector while the low-energy toroidal states are basically isoscalar).

It is also interesting that the lowest toroidal 7.92-MeV state demonstrates
a strong collective $0^+0 \to 3^-1$ transition with $B(E31)$ = 12 W.u..
This means that, though 7.92-MeV state is mainly vortical, it also has some
irrotational octupole component. Appearance of this component
is explained by the large axial quadrupole deformation in
$^{24}$Mg, which leads to the strong mixing of the dipole and octupole
modes. In the 7.92-MeV state, the octupole irrotational
fraction seems to dominate over the dipole irrotational one.
  Note also that 2qp configurations $211\uparrow-330\uparrow$
dominating in 7.92-MeV and 9.79-MeV states
(see Table I) fulfill  the asymptotic selection rules for E31 and M21
transitions \cite{BM59,Sol}
(E31: $\Delta N=\pm1, \pm3$, $\Delta n_z = 0, \pm2$, $\Delta \Lambda = 1$;
 M21: $\Delta N=\pm1, \pm3$, $\Delta n_z = 0, \pm1, \pm2$, $\Delta \Lambda = 0, 1$),
 and not for E11 ($\Delta N = \pm1$, $\Delta n_z = \pm0$, $\Delta \Lambda = 0$).
 This favors E31 and M21 transitions but hinders E11 ones. In 9.79-MeV state,
 $B(E31)$ is small because of the mutual compensation of proton and neutron contributions.
 In  7.92-MeV state, the hindered $B(E11)= 10^{-3}-10^{-4}$ W.u. is nevertheless
 essentially larger  than the experimental value 3.3 $10^{-6}$ W.u.
 \cite{NDS24Mg}. So perhaps the irrotational dipole component in this state is
weaker than in our calculations.

The left part of Fig.~\ref{fig-4} shows
transition probabilities for K=0 states. The  9.56-MeV
state has hindered E10 and enhanced E30 strengths (see also Table II).
In this state, the signature $\gamma$ coincides with the parity ($\gamma =\pi$ = - 1),
its rotational band is $I^{\pi}=1^-, 3^-, ...$, and so the magnetic decay
to the ground state is absent. Instead we have a noticeable amount of
$B(M20, 0^+0_{\rm gs}  \to 2^-0)$ strength (together with vanishing $B(E10)$ and $B(E30)$)
for the higher state at 9.93 MeV with $\gamma =-\pi$ = + 1.
This state is not toroidal and so out of our interest. At the same time, this example
shows that non-toroidal states  can also have significant $B(M2)$.
Thus, a large $B(M2)$ may be used for discrimination of the toroidal mode only in
low-energy states with K=1.

The experimental data for low-energy spectra in $^{24}$Mg \cite{NDS24Mg}
show $1^-$ levels at 7.555 and 8.437 MeV. Both levels can be reasonable candidates
for toroidal excitations \cite{Ne_PRL18}. Moreover, the direct decay
(most probably M2) from the first $I^{\pi}=2^-$ state at 8.864 MeV  to the
ground state is observed \cite{NDS24Mg}. The decay is weak as compared
with other decay channels of this state. Our QRPA approach is not
enough to describe the complicated decay scheme in $^{24}$Mg.
Nevertheless it allows to state that orbital
M21 transitions from low-energy K=1 states can serve as
promising indicators of the toroidal mode in deformed nuclei.
\begin{figure} 
\includegraphics[width=8.5cm,clip]{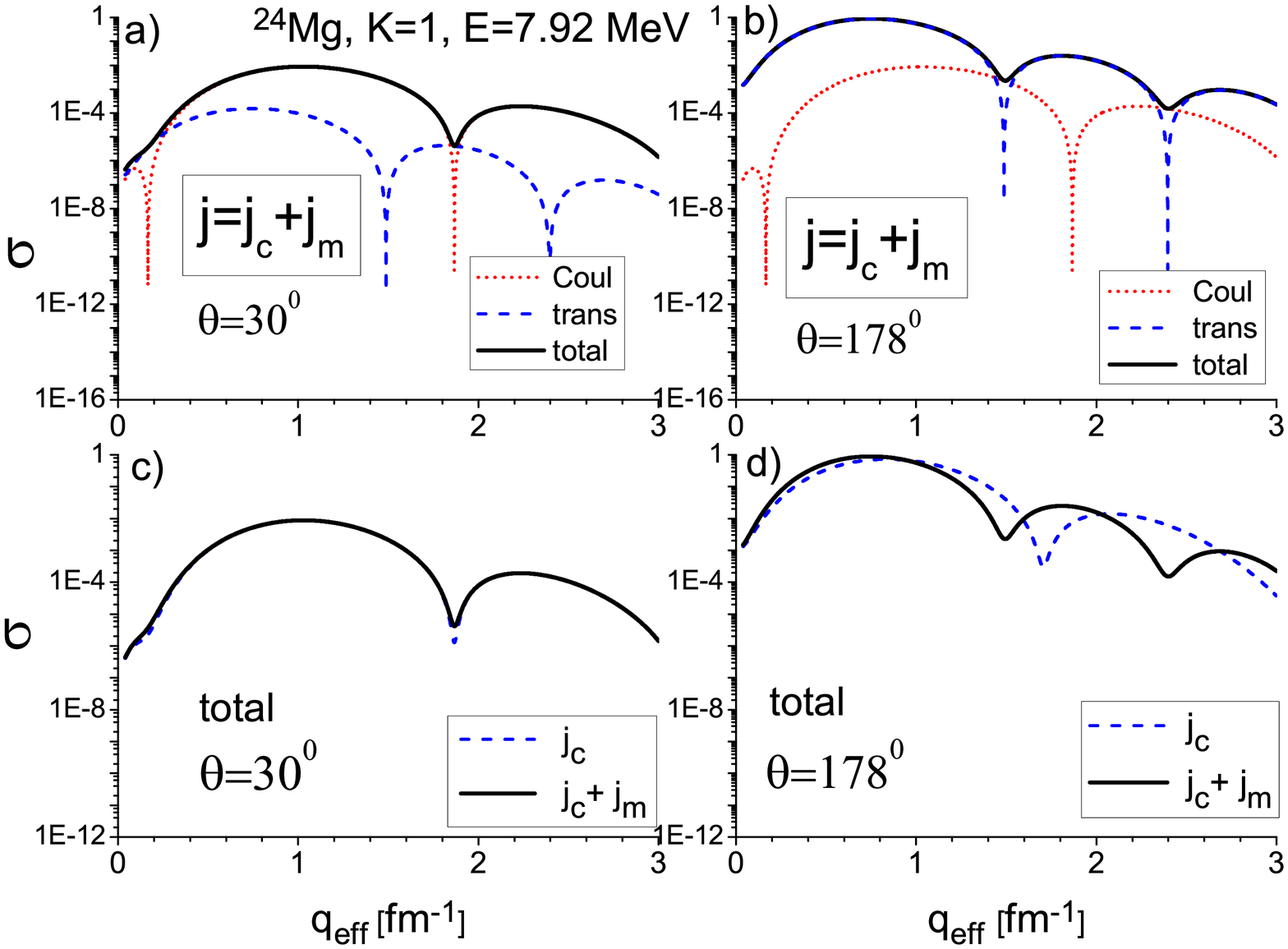}
\caption{The $(e,e')$ cross-section
for $I^{\pi}K=1^-1$  state at 7.92 MeV, calculated for the scattering
angles $\theta = 30^{\circ}$
(left)  and $178^{\circ}$ (right). In the upper plots (a)-(b), the Coulomb, transversal
and total cross-sections are compared. In the bottom plots (c)-(d), the total
cross-sections with and without ${\bold j}_{\rm m}$ are shown.
}
\label{fig-7}       
\end{figure}

\subsection{$(e,e')$ reaction}

To discriminate  TS from other dipole modes, we need a reaction sensitive
to the nuclear interior.  The inelastic electron scattering $(e,e')$
is just the proper case.
In the Plane Wave Born Approximation (PWBA), the $(e,e')$ cross-section
for $E(M)\lambda$ excitations reads
\cite{HB83}
\begin{eqnarray}
\label{PWBAcs}
&&\frac{d\sigma}{d\Omega} (\theta, q, E_i) =
4\pi \sigma_{\rm Mott}(\theta, E_i) f_{\rm rec}(\theta, E_i)
\\
&\cdot& \left[[F^C_{E\lambda} (q)]^2
+ \left(\frac{1}{2}+\tan^2(\frac{\theta}{2})\right)
\left( [F^T_{M\lambda} (q)]^2 + [F^T_{E\lambda} (q)]^2 \right) \right]
\nonumber
\end{eqnarray}
where $\sigma_{\rm Mott} (\theta, E_i)$ is the Mott cross section for the unit charge,
$f_{\rm rec}(\theta, E_i)$ is the recoil factor, $E_i$ is the incident electron energy,
$\theta$ is the scattering angle. Further,  $F^C_{E\lambda}(q)$, $F^T_{E\lambda}(q)$, and
$F^T_{M\lambda}(q)$ are Coulomb and transversal electric and  magnetic form factors as
a function of the  momentum transfer $q$.
Here, $q=(2/(\hbar c)\sqrt{E_i E_f} \sin (\theta/2)$ where $E_f=E_i-E_{\nu}$ is the final
electron energy and $E_{\nu}$ is the nuclear excitation energy.
For the light nucleus $^{24}$Mg, the Coulomb distortions should be small and so
PWBA is the relevant approximation. We also can use
$f_{\rm rec}(\theta, E_i)$=1. To take roughly into account  the Coulomb distortions,
the figures below are plotted  as a function of the effective momentum
transfer
\begin{equation}
q_{\rm eff} = q(1+1.5 \frac{Z \alpha \hbar c}{E_i R})
\end{equation}
where $Z$ is the nuclear charge and $R=1.12 A^{1/3}$ fm.

First of all, let's consider the $(e,e')$ cross section for the toroidal
states and inspect the effect of the magnetization current ${\bold j}_{\rm m}$ on
them. Since the toroidal mode is transversal \cite{Bas93,Mis06,Dub75},
it is natural to look for its signature in the dipole transversal electric
form factor $F^T_{E1}$ at backward scattering angles.

In Fig.~\ref{fig-7}, the normalized cross-section
$\sigma = \frac{d\sigma}{d\Omega}/\sigma_{\rm Mott}$
for the 7.92-MeV state in $^{24}$Mg is  plotted for small $\theta$ = 30$^{\circ}$ and
large $\theta$ = 178$^{\circ}$ scattering angles. Panels (a)-(b) show that, as expected,
the total cross-section is dominated by the Coulomb part at $\theta$ = 30$^{\circ}$ and by
electric transversal part at $\theta$ = 178$^{\circ}$. Further, panels (c)-(d) show that inclusion of
$j_{\rm m}$ does not almost influence the cross-section at $\theta$ = 30$^{\circ}$  but leads
to considerable changes for $q_{\rm eff} > 1$ fm$^{-1}$ at the backward angle $\theta = 178^{\circ}$.
The latter significantly complicates the direct search of TS
in the transversal cross-section at large $\theta$.
\begin{figure} 
\includegraphics[width=8.5cm,clip]{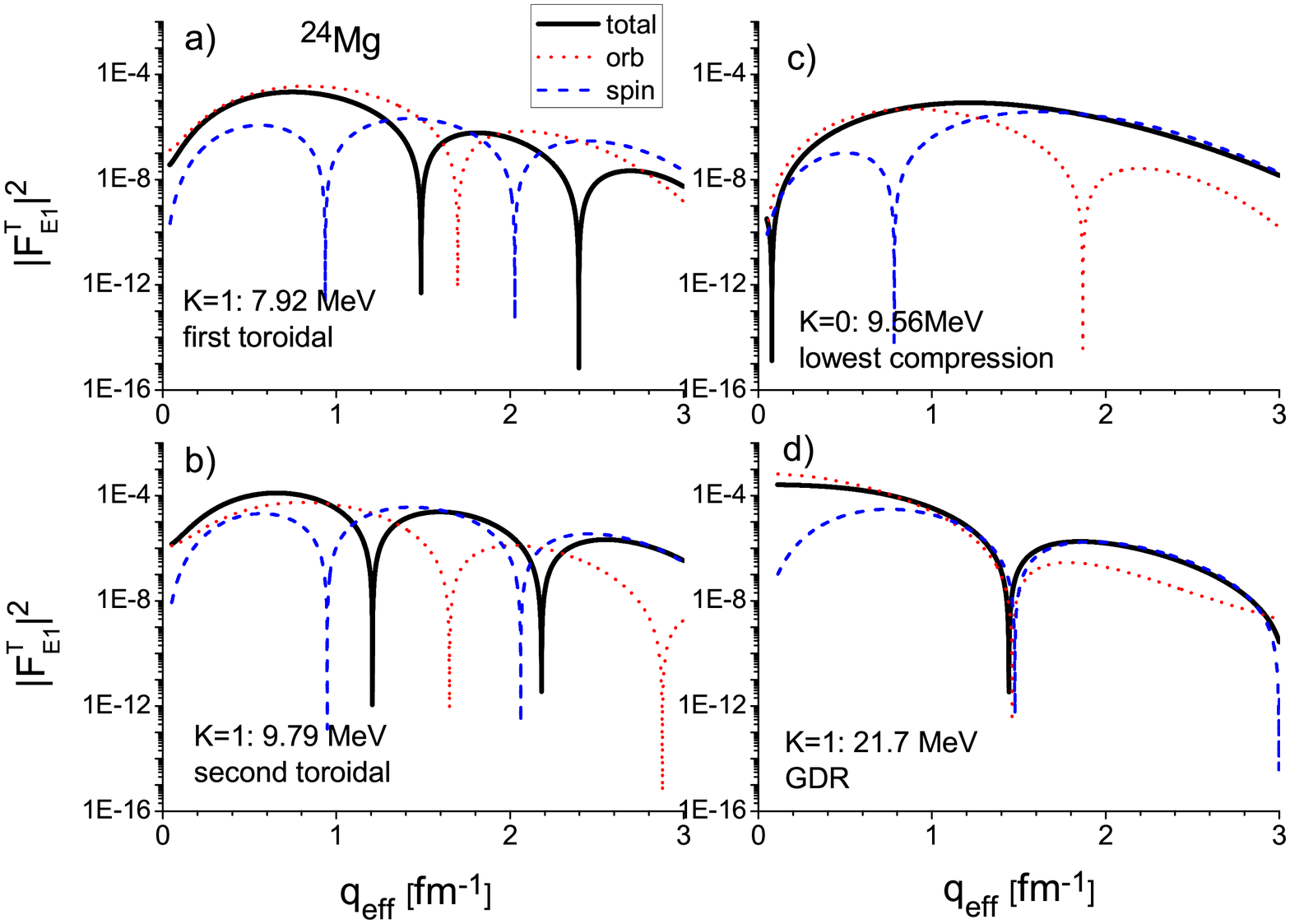}
\caption{Squared electric transversal form factors $|F^T_{E1}|^2$
 calculated with the total (black solid line), orbital (red dotted line)
and spin (dash blue line) nuclear current
for different QRPA states: toroidal $I^{\pi}K=1^-1$ at 7.92 MeV (a) and
9.79 MeV (b),
compressional $1^-0$  at 9.56 MeV (c), and
GDR $1^-1$ at 21.7 MeV (d).}
\label{fig-8}       
\end{figure}
\begin{figure} 
\includegraphics[width=8.5cm,clip]{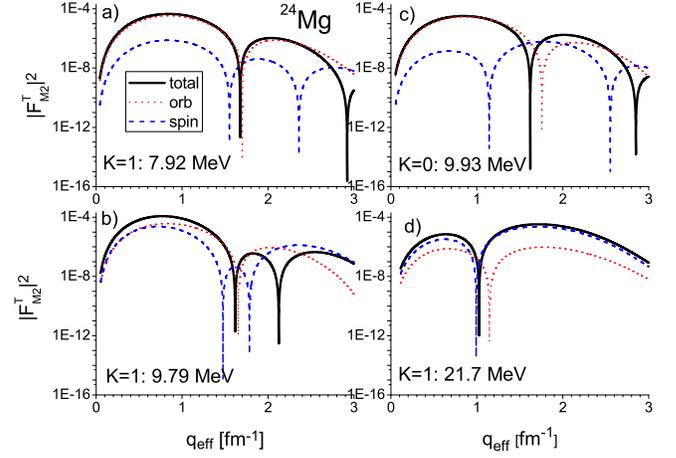}
\vspace{2mm}
\caption{Squared magnetic  transversal form factors $|F^T_{M2}|^2$
 calculated with the total (black solid line), orbital (red dotted line)
and spin (dash blue line) nuclear current
for different QRPA states: toroidal $I^{\pi}K=1^-1$ at 7.92 MeV (a) and
9.79 MeV (b),
compressional $1^-0$  at 9.93 MeV (c), and
GDR $1^-1$ at 21.7 MeV (d).}
\label{fig-9}       
\end{figure}

We see that the Coulomb cross-section for the  toroidal 7.92-MeV state
has a distinctive minimum at $q_{\rm eff} < 0.2 \; \rm{fm}^{-1}$. Similar
minima were earlier found
for low-energy dipole states in light N=Z spherical doubly magic nuclei like
$^{16}$O, see \cite{Mi75} for experiment and  \cite{Ca90,Papa11,Ne_20Ne} for discussion.
Following \cite{Papa11}, these states can also exhibit toroidal flow.
Most probably, however, these minima are caused not by toroidal flow but rather by
destructive competition between the dominant T=0 and minor T=1 components in these states
\cite{Ca90,Papa11}.

Nevertheless the toroidal mode leaves in $(e,e')$ scattering some signatures suitable
for its discrimination. These signatures are illustrated in
Fig.~\ref{fig-8} where the squared transversal form factors
$|F^T_{E1}|^2$ for different dipole states in $^{24}$Mg are plotted. Here we consider
the toroidal K=1 states at 7.92 and 9.79 MeV, the compressional K=0 state
at 9.56 MeV and the high-energy K=1 state at 21.7 MeV from the isovector giant dipole
resonance (GDR). The form factors are calculated with the total
${\bold j}_{\rm c}+{\bold j}_{\rm m}$, convective (orbital) ${\bold j}_{\rm c}$,
and spin ${\bold j}_{\rm m}$ nuclear currents.


Fig.~\ref{fig-8} shows that total form factors for toroidal 7.92-MeV and 9.79-MeV
states (plots (a)-(b)) are more structured as they have two diffraction minima
at $q_{\rm eff} < 3 \; {\rm fm}^{-1}$) and, in this sense, significantly
deviate from the form factors for other states (plots (c)-(d)). In the orbital
form factors, the diffraction minima lie essentially higher than in the spin ones.
For toroidal states, just
destructive interference of the orbital and spin contributions gives
diffraction minima in the total $|F^T_{E1}|^2$. They are at $q^{\rm min}_{\rm eff} = 1.50,
 2.39 \; \rm{fm}^{-1}$ in 7.92-MeV state and at $q^{\rm min}_{\rm eff} = 1.20,
2.17 \; \rm{fm}^{-1}$ in 9.79-MeV state. Neither orbital, nor spin
contribution alone can describe the behavior of the total $|F^T_{E1}|^2$.
Therefore this behavior can be used  as a sensitive tool
for determination of the orbital/spin interplay. One may state that
the QRPA $\nu$-th wave function
correctly describes  the orbital and spin contributions only if
it reproduces the features of the total $|F^T_{E1}|^2$ at large scattering angles.

Note also that, following panels (a)-(b),
the orbital (toroidal) contribution dominates
over the spin one at $q_{\rm eff} < 1.1 \; \rm{fm}^{-1}$.
The dominance is impressive for 7.92-MeV state.

Further, Fig.~\ref{fig-9} shows the squared magnetic form factors
$|F^T_{M2}|^2$ calculated with the total ${\bold j}_{\rm c}+{\bold j}_{\rm m}$,
convective (orbital)  ${\bold j}_{\rm c}$, and spin ${\bold j}_{\rm m}$
nuclear currents. In the plots (a), (b), and (d), excitations $I^{\pi}K=2^-1$ related to
the states in Fig.~\ref{fig-8} are considered. In the plot (c),
we consider the  compressional K=0 9.93-MeV state with the signature $\gamma =-\pi$ = + 1 and non-zero
$B(M20, 0^+0_{\rm gs} \to 2^-0)$ value (see Table II).
Fig.~\ref{fig-9} (a) shows that, in the toroidal K=1 7.92-MeV state, the orbital contribution
strongly dominates at $q_{\rm eff} < 1.6 \; \rm{fm}^{-1}$.
The same takes place  in Fig.~\ref{fig-9} (c) for K=0 9.93-MeV state.
In both cases, the first diffraction minimum is fully determined  by the orbital form
factor. In the toroidal K=1 9.79-MeV state,
the dominance of the orbital part is weaker but we have the specific second
diffraction minimum at $q^{\rm min}_{\rm eff} = 2.12 \; \rm{fm}^{-1}$, produced by the
destructive interference of the orbital and spin parts. For both toroidal 7.92-MeV and
9.79-MeV states, the formfactors  $|F^T_{M2}|^2$
are structured enough to probe the wave functions of these states
and judge on the important (or even dominant) role of the orbital flow.

Altogether one may propose the following two-step scheme for discrimination of
individual vortical toroidal states in $(e, e')$ reaction.

1) The calculations (e.g. QRPA) should identify the relevant candidates
for the toroidal dipole states. They should be low-energy K=1 states
with the following properties: i) large toroidal strength like in
Figs.~\ref{fig-1} and \ref{fig-4},
ii) typical toroidal picture for the convective current density, like
in Figs.~\ref{fig-2} and \ref{fig-5}, iii) enhanced $B(M2)$ and weak
$B(E1)$ transition rates for decays to the ground state.

2) The wave functions of the chosen states should reproduce the main features
of the total squared transversal form factors  $|F^T_{E1}|^2$ and $|F^T_{M2}|^2$
at back scattering angles. In particular, magnitudes of form factors at diffraction maxima
and positions of diffraction minima should be described.
As shown in our study, the behavior of these form-factors is very
sensitive to the interference of the spin and orbital contributions.
If the experimental  $(e,e')$  data are not reproduced by the spin contribution alone
but reasonably described by the total spin + orbital contribution, then: i) wave functions
of the chosen states can be assumed as reliable and ii) toroidal distributions of their
convective currents can be considered as realistic.

To check this two-step scheme,
the new $(e,e')$  experiments for $^{24}$Mg are desirable. For this aim, the electron beams
with the incident electron energy 40-90 MeV, available e.g. in Darmstadt facilities
\cite{PVNC99,Ri04}, could be used.

Note that a similar prescription was earlier employed for confirmation
of the vortical twist M2 mode in Darmstadt $(e,e')$ experiment \cite{PVNC99}. Namely,
the orbital M2 contribution to the backward electron scattering
was justified by comparison of the calculated spin and spin+orbital M2 form factors
with experimental data. The fact that only spin + orbital contribution
(but not spin contribution alone) was sufficient to describe the experimental data,
was claimed as a robust signal of a strong orbital twist M2 flow.

Note that our QRPA calculations do not take into account such factors as
the nuclear triaxiality and coupling with complex configurations (CCC).
By our opinion, these factors should not essentially change our main results. Indeed,
following various calculations \cite{Rod10,Be08,Yao11,Hino11,Kimura12},
$^{24}$Mg has a weak triaxial
softness  in the ground state and more triaxiality in positive-parity excited states.
In the lowest negative-parity dipole states, the triaxiality is found {\it negligible} in
K=1 and significant in K=0 excitations \cite{Kimura12}. Since we mainly address
low-energy K=1 excitations with the dominant large-magnitude axial prolate deformation,
the treatment of $^{24}$Mg as an axial prolate nucleus should be reasonable. Besides,
the dipole K=1 states of our interest have a low collectivity and so should exhibit a minor
CCC impact.

\section{Conclusions}

A possibility to search individual E1 toroidal states (TS) in inelastic
electron scattering $(e,e')$ to back angles was scrutinized within the
self-consistent quasiparticle random-phase-approximation (QRPA) model using
the Skyrme force SLy6. As a relevant example, the low-energy dipole states
with K=0 and 1 in axially deformed $^{24}$Mg were thoroughly explored.
We inspected vortical toroidal and irrotational compressional E1 responses,
transition rates $B(E1, 0^+0_{\rm gs}  \to 1^-K),
\; B(E3, 0^+0_{\rm gs}  \to 3^-K)$ and $B(M2, 0^+0_{\rm gs}  \to 2^-K)$,
distributions of transition currents, form factors and cross sections for
$(e,e')$ reaction. The cross sections were calculated in the Plane Wave
Born Approximation. In the relevant cases, the separate contributions of the
convection  ${\bold j}_{\rm c}$, and magnetization ${\bold j}_{\rm m}$
parts of nuclear current were analyzed.

The analysis of these results led to a two-step scheme for the search
of toroidal K=1 states in $(e,e')$ scattering. In the {\it first} step,
QRPA calculation are used to determine the promising candidates for
toroidal states (with large toroidal responses, distinctive
toroidal distribution of the convective nuclear current and significant $B(M2)$ values).
In the {\it second} step, these states are checked to reproduce the pattern of the
experimental data
for E1 and M2 transversal form factors in $(e,e')$ scattering to back angles.
Following our analysis, these form factors exhibit a strong interference
of the convective (orbital) and magnetization (spin) contributions of the nuclear
current and this interference determines, in a large extent,
the features (form factor maxima, positions of the first two diffraction minima, etc)
of the form factors. As a result, E1 and M2 transversal form factors
can serve as sensitive probes for the interplay between orbital and spin
contributions. If only the combined spin+orbital contributions (but not spin alone)
allow to reproduce the experimental behavior of  these form factors, then one may claim
that i) the structure of the chosen calculated state correctly matches the orbital and
spin fractions  and ii) the toroidal  distribution of the nuclear current in this state is indeed
realistic. A similar prescription was earlier used  in the experimental
search of the vortical twist M2 mode in $(e,e')$ reaction \cite{PVNC99}. Note that
involvement of $B(M2)$ values and M2 form factors for discrimination of E1 toroidal states
is relevant only for deformed nuclei and this part of the analysis
should be skipped in spherical nuclei.

In the proposed identification scheme, the interference between the orbit and spin contributions
to the experimentally accessible  $(e,e')$ form factors is the key element.
The toroidal strengths and current distributions as such
can hardly be measured directly, but can be used as pre-selectors
to choose from QRPA calculations the proper candidate states
 for the further analysis of $(e,e')$ scattering.

In the present study for $^{24}$Mg, two individual toroidal K=1 states at 7.92 and 9.97 MeV
were found and thoroughly explored. It was shown that the magnetization current
${\bold j}_{\rm m}$ has a strong impact for these states.
Just this considerable magnetic contribution together with the dominant orbital contribution
leads to the significant interference effects in E1 and M2 form factors and
so paves the way for discrimination of the toroidal states. Furthermore, we have found
that ${\bold j}_{\rm m}$ can produce itself the magnetization toroidal states.

The above scheme can be also used for heavier nuclei where
we deal not with individual toroidal states but rather with broadly spread toroidal
strength functions. In this case, we should work with the averaged characteristics
using the technique described in Ref. \cite{Rep13} for $^{208}$Pb.

In principle, similar schemes can be applied to other reactions
($(e, e'\gamma)$, $\alpha, \alpha')$, $(p,p')$, etc) relevant for the search of
toroidal dipole states (see Ref. \cite{Bracco19} for the recent review
of various reactions for dipole excitations).
By our opinion, there is no
problem to excite TS in nuclei. Following our present analysis for $^{24}$Mg and
previous analysis for a variety of medium and heavy nuclei \cite{Kv11,Rep17,RNKR_EPJA},
even basically vortical dipole states usually have a minor irrotational
fraction \cite{RNKR_EPJA} and this fraction can be used as a {\it doorway}
for excitation of the toroidal mode in various reactions. The main trouble
is not to excite vortical TS but to identify them.
This is a part of a general fundamental problem of identification of vorticity in nuclei.
The problem is indeed demanding since its solution requires a theory-assisted
analysis combining information on nuclear structure and reaction mechanisms.
Hopefully, the search of the vortical toroidal mode in $(e,e')$ reaction
will be an important and encouraging  step in this direction.

\section*{Acknowledgement}

V.O.N. thanks Profs. P. von Neumann-Cosel, J. Wambach and V.Yu. Ponomarev
for useful discussions. The work was partly supported by
Heisenberg - Landau (Germany - BLTP JINR) and Votruba - Blokhintsev
(Czech Republic - BLTP JINR) grants. A.R. is grateful for support
from Slovak Research and Development Agency (Contract
No. APVV-15-0225). J.K. thanks the grant of Czech Science Agency
(Project No. 19-14048S).

\end{document}